\documentclass[aps,prb,superscriptaddress,twocolumn,showpacs,preprintnumbers,amsmath,amssymb]{revtex4-2}

\usepackage{graphicx}
\usepackage{color}
\usepackage{bm}
\begin{document}

%\preprint{APS/123-QED}

\title{Observation of two-level critical-state in a van-der-Waals superconductor Pt(Bi$_{1-x}$Se$_x$)$_2$}% Force line breaks with \\

\author{Y.~Samukawa}
\affiliation{Department of Physics, Graduate School of Science, Osaka University, Toyonaka 560-0043, Japan}
\author{M.~Maeda}
\affiliation{Department of Physics, Graduate School of Science, Osaka University, Toyonaka 560-0043, Japan}
\author{N.~Jiang}
\email{nan.jiang@phys.sci.osaka-u.ac.jp}
\affiliation{Department of Physics, Graduate School of Science, Osaka University, Toyonaka 560-0043, Japan}
\affiliation{Center for Spintronics Research Network, Osaka University, Toyonaka 560-8531, Japan}
\affiliation{Institute for Open and Transdisciplinary Research Initiatives, Osaka University, suita 565-0871, Japan}
\author{R.~Nakamura}
\affiliation{Department of Physics, Graduate School of Science, Osaka University, Toyonaka 560-0043, Japan}
\author{M.~Watanabe}
\affiliation{Department of Physics, Graduate School of Science, Osaka University, Toyonaka 560-0043, Japan}
\author{K.~Takaki}
\affiliation{Department of Physics, Graduate School of Science, Osaka University, Toyonaka 560-0043, Japan}
\author{Y.~Moriyasu}
\affiliation{Department of Physics, Graduate School of Science, Osaka University, Toyonaka 560-0043, Japan}
\author{K.~Kudo}
\affiliation{Department of Physics, Graduate School of Science, Osaka University, Toyonaka 560-0043, Japan}
\affiliation{Institute for Open and Transdisciplinary Research Initiatives, Osaka University, suita 565-0871, Japan}
\author{Y.~Niimi}
\affiliation{Department of Physics, Graduate School of Science, Osaka University, Toyonaka 560-0043, Japan}
\affiliation{Center for Spintronics Research Network, Osaka University, Toyonaka 560-8531, Japan}
\affiliation{Institute for Open and Transdisciplinary Research Initiatives, Osaka University, suita 565-0871, Japan}

\date{\today}% It is always \today, today,

\begin{abstract}
Trigonal PtBi$_2$ is one of the attractive van-der-Waals materials because of the enhancement of its superconducting transition temperature $T_{\rm{c}}$ by doping chalcogen elements such as Se and Te. Recently, it has been reported that $T_{\rm{c}}$ of Pt(Bi$_{1-x}$Se$_x$)$_2$ is enhanced by a factor of 4, compared to the pristine PtBi$_2$, together with the polar-nonpolar structural phase transition. Thus, it is desirable to study electrical transport properties for this new superconducting compound. Here, we have performed magnetotransport measurements for Pt(Bi$_{1-x}$Se$_x$)$_2$ ($x$ = 0.06 and 0.08) thin-film devices and have observed a peculiar magnetoresistance where a finite hysteresis appears when the superconducting state is broken. By measuring the magnetoresistance systematically, we have attributed this magnetoresistance to the two-level critical-state where fluxons pinned in Pt(Bi$_{1-x}$Se$_x$)$_2$ play an important role.
\end{abstract}

\maketitle

%\tableofcontents

\section{\label{sec:level1}introduction}
In recent years, van-der-Waals (vdW) materials have attracted much attention from the viewpoints of the low-dimensional physics as well as of the next generation electronic devices due to easy processing into thin-film forms.
Amongst a wide variety of vdW materials, ultra-thin transition metal dichalcogenides such as MoS$_2$~\cite{mos2_1,mos2_2,mos2_3,mos2_4}, NbSe$_2$~\cite{nbse2_1,nbse2_2,nbse2_3}, TaS$_2$~\cite{tas2_1,tas2_2,tas2_3,tas2_4}, and WTe$_2$~\cite{wte2_1,wte2_2} are being studied extensively from the perspective of two-dimensional superconductivity. 
In order to expand the material groups of thin-film superconductors, it is important to explore other vdW superconductors.

Trigonal PtBi$_2$ is an attractive noncentrosymmetric vdW material 
with a space group of $P31m$ 
[see Fig.~1(a)]~\cite{ptbi2_1}. 
It has fascinating properties such as huge linear magnetoresistance due to the band structure~\cite{ptbi2_lmr1,ptbi2_lmr2}, large Rashba splitting~\cite{ptbi2rashba}, superconductivity at the transition temperature $T_{\rm{c}}$ = 0.6 K~\cite{ptbi2}, pressure-induced superconductivity~\cite{PtBi2Press}, and triple point near the Fermi surface~\cite{TP}. 
By doping Se into the Bi site, the polar-nonpolar structural phase transition takes place at the doping rate $x$ of 3.2\%, showing the centrosymmetric crystal structure with a space group $P\bar{3}m1$ [see Fig.~1(b)]~\cite{takaki}. 
Along with the structural transition, $T_{\rm{c}}$ increases from 0.6~K to 2.4~K. 
After this critical doping rate, $T_{\rm{c}}$ monotonically decreases. Although the microscopic origin of the enhancement of $T_{\rm{c}}$ is not clear, Pt(Bi$_{1-x}$Se$_x$)$_2$ is a new type of vdW superconductor in which $T_{\rm{c}}$ increases with structural phase transition. 

In this paper, we have studied electrical transport properties in superconducting Pt(Bi$_{1-x}$Se$_x$)$_2$ thin-film devices with $x$ = 0.06 and 0.08. A clear hysteresis in magnetoresistance was observed near the critical magnetic field $\mu_{0}H_{\rm{c2}}$, where $\mu_{0}$ is the permeability in vacuum. This hysteresis has the opposite sign to the conventional hysteresis: in other words, the superconductivity is broken with a small applied magnetic field in comparison to the applied field when the superconductivity is recovered. By measuring $\mu_{0}H_{\rm{c2}}$ not only as a function of temperature but also as a function of magnetic-field sweep rate and maximum applied field, we have attributed this hysteresis to the two-level critical-state caused by inhomogeneities of the superconducting state~\cite{tinkham}.

\section{\label{sec:level1}experimental methods}
Single crystalline samples of Pt(Bi$_{1-x}$Se$_x$)$_2$ were synthesized by using the same method as a previous study reported by some of the present authors~\cite{takaki}. A stoichiometric mixture of Pt (99.95\%), Bi (99.99\%), and Se (99.9\%) powders was sealed in an evacuated quartz tube and heated at 630~$^{\circ}$C for 24 hours, followed by quenching in ice water. The chemical compositions of the obtained samples were estimated by energy dispersive X-ray spectroscopy (EDS) using a TM4000Plus II scanning electron microscope (Hitachi High-Tech) equipped with an AztecOne energy dispersive spectrometer (Oxford Instruments). The crystal structure was examined by powder X-ray diffraction (XRD) using a MiniFlex600-C X-ray diffractometer equipped with a D/teX Ultra2-high-speed one-dimensional detector (Rigaku). We obtained thin flakes through the mechanical exfoliation technique using scotch tapes. The flakes were transferred onto Si/SiO$_2$ substrates, and the electrodes were patterned by electron beam lithography, followed by depositing Ti/Au films. An optical microscope image of our typical device is shown in the inset of Fig.~1(d). The thickness $d$ of each flake was measured by an atomic force microscope (NanoNaviReal Probe Station). The devices were cooled using a dilution refrigerator (Oxford Instruments) with a superconducting magnet or a $^4$He flow refrigerator with an electromagnet, depending on the temperature range to be measured. The measurements of electrical transport properties were performed by the conventional four terminal method through a lock-in amplifier (SR-830).

\begin{figure}
\includegraphics[width=85mm]{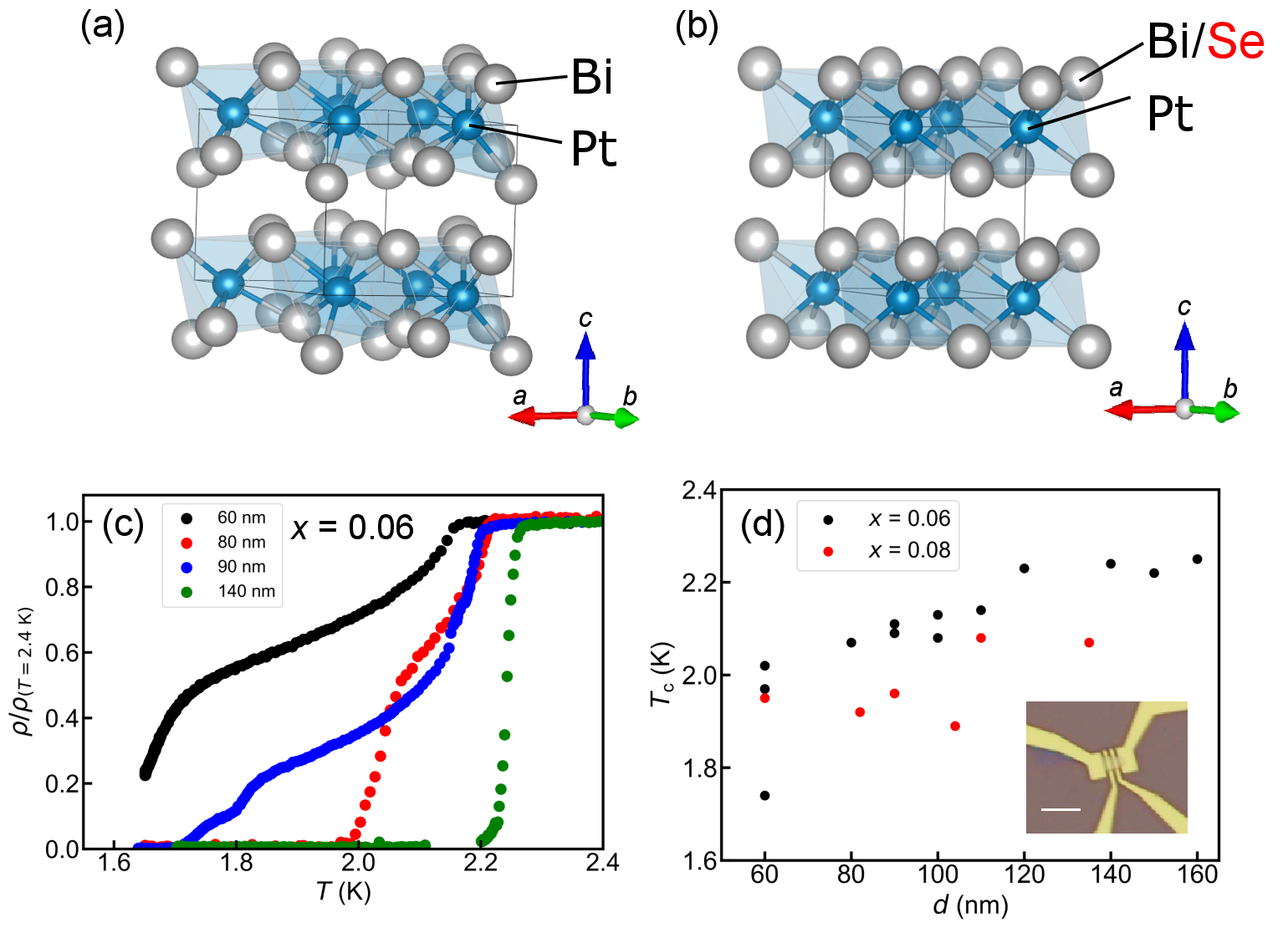}
\caption{\label{fig:epsart}(a), (b) Crystal structures of PtBi$_2$ with a space group of $P31m$ and Pt(Bi$_{1-x}$Se$_x$)$_2$ with a space group of $P\bar{3}m1$, respectively. (c) Temperature dependence of electrical resistivity $\rho$ near $T_{\rm{c}}$ for $x$ = 0.06 with different thicknesses, 60 nm (black dots), 80 nm (red dots), 90 nm (blue dots), and 140 nm (green dots). The vertical axis is normalized by $\rho$ at $T=2.4$~K. (d) Thickness $d$ dependence of $T_{\rm{c}}$ for $x$ = 0.06 (black dots) and $x$ = 0.08 (red dots) devices. The inset is an optical microscope image of a typical device. The white bar in the inset corresponds to 5 $\mu$m.}
\end{figure}

\section{\label{sec:level1}results and discussions}
In Fig. 1(c), we show the temperature dependence of electrical resistivity $\rho$ normalized by the resistivity at 2.4~K, $\rho (T=2.4~\rm{K})$, for $x = 0.06$ devices with several different thicknesses. The resistivity starts to drop at around 2.2~K, reflecting the onset of the superconductivity phase transition. The temperature where $\rho$ reaches zero is lower for thinner devices. To characterize such a tendency, we plot $d$ dependence of $T_{\rm{c}}$, defined by half the resistivity of the normal state, for $x = 0.06$ and 0.08 in Fig.~1(d). $T_{\rm{c}}$ decreases with decreasing $d$ for both Se concentrations. This tendency is consistent with other thin-film superconductors~\cite{thickness1,thickness2,thickness3} .

This $d$ dependence of $T_{\rm{c}}$ can be explained by the inhomogeneity in the superconducting state. In Pt(Bi$_{1-x}$Se$_x$)$_2$, the superconductivity should be non-uniform: in other words, there should be strong (or weak) superconducting regions. Such an inhomogeneity should be induced by spatially non-uniform doping of Se and spatially non-uniform strain. When $d$ approaches the characteristic inhomogeneity length, $T_{\rm{c}}$ would be reduced as discussed in a previous study~\cite{fetese}. The inhomogeneity of the superconductivity also leads to the broadening of the superconducting transition in the $\rho$ versus $T$ curve in Fig.~1(c). We will discuss this inhomogeneity in the last part of this section.

\begin{figure}
\includegraphics[width=85mm]{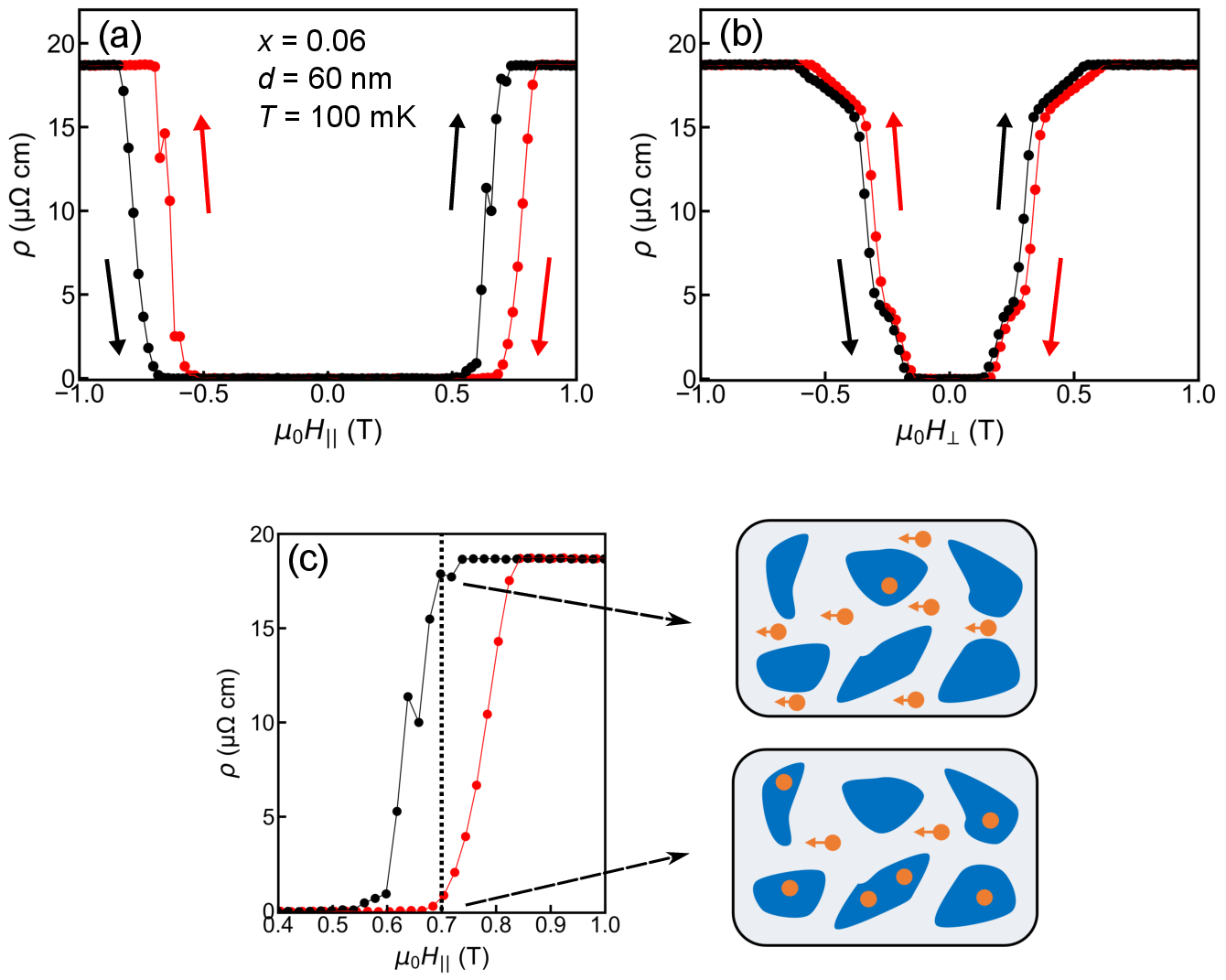}
\caption{\label{fig:epsart} (a), (b) Magnetic field dependence of $\rho$ under (a) the in-plane and (b) out-of-plane magnetic fields measured with a device of $x = 0.06$ and $d= 60$~nm at 100~mK. The magnetic-field sweep rate is 0.065~T/min and the applied maximum field is 2~T. Black and red dots represent data from $-2$~T to 2~T and data from 2~T to $-2$~T, respectively. We only show the magnetic field range in between $-1$~T and $1$~T to focus on the hysteresis behavior. (c) Closeup view of (a), together with schematic images of fluxons trapped inside the device based on the two-level critical-state model. The blue and gray areas in the schematic images represent strong and weak superconductivity regions, respectively. The orange dots represent fluxons in the superconductivity region. The fluxons in the weak superconductivity region can move by the electric current and generate a voltage, while the fluxons in the strong superconductivity region cannot move by the strong pinning potential, resulting in zero resistivity.}
\end{figure}

Next, we discuss magnetotransport properties of Pt(Bi$_{1-x}$Se$_x$)$_2$ thin-film devices. In Figs.~2(a) and 2(b), we show the in-plane magnetic field $\mu_{0}H_{||}$ and out-of-plane magnetic field $\mu_{0}H_{\perp}$ dependences of $\rho$ at 100 mK measured with a device of $x = 0.06$ and $d=60$~nm. We have observed a hysteresis near $\mu_{0}H\rm{_{c2}}$ in both cases. We note that the observed hysteresis is opposite to the well-known hysteresis.
When the magnetic field is swept from zero to a large value (positive sweep), $\mu_{0}H\rm{_{c2}}$ is smaller compared to the case when the magnetic field is swept from a large value to zero (negative sweep) in our devices as shown in Figs~2(a) and 2(b), while $\mu_{0}H\rm{_{c2}}$ of the positive sweep is larger than that of the negative sweep in the conventional hysteresis. 
The magnitude of the hysteresis is larger for the in-plane configuration compared to the out-of-plane configuration.

\begin{figure}
\includegraphics[width=60mm]{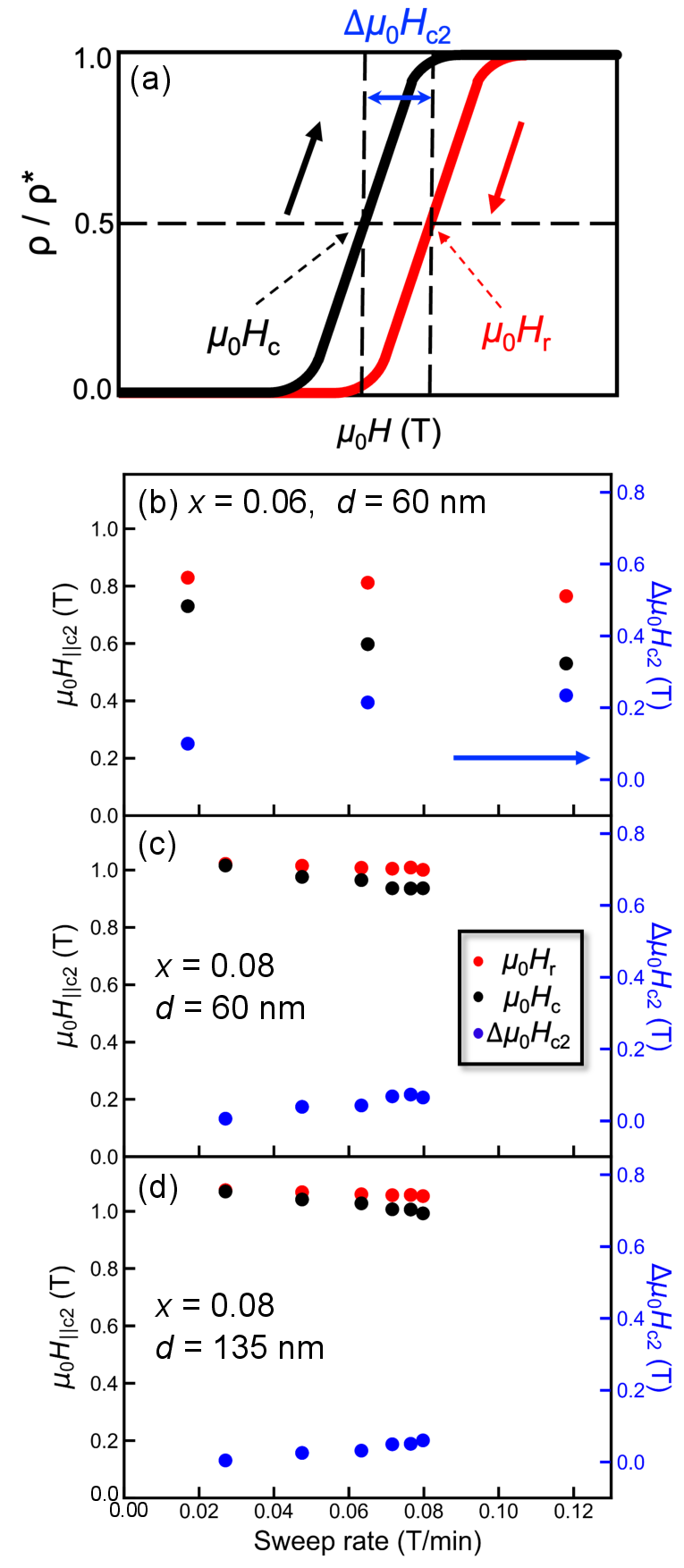}
\caption{\label{fig:epsart} (a) A schematic illustration of the hysteresis behavior to define each parameter. The vertical axis is normalized by $\rho$ at $\mu_{0}H=1.2$~T ($\equiv\rho^*$). $\mu_{0}H\rm{_{r}}$ and $\mu_{0}H\rm{_{c}}$ are defined as the magnetic fields at which $\rho$ becomes half of the resistivity in the normal state during the negative and positive sweeps, respectively. The magnitude of the hysteresis is defined as $\Delta\mu_{0}H\rm{_{c2}} \equiv \mu_{0}$$H\rm{_{r}} - \mu_{0}$$H\rm{_{c}}$. (b)-(d) The magnetic-field sweep rate dependence of $\mu_{0}H\rm{_{r}}$ (red dots), $\mu_{0}H\rm{_{c}}$ (black dots), and $\Delta\mu_{0}H\rm{_{c2}}$ (blue dots) measured at $T = 100$~mK with devices of (b) $x = 0.06$ and $d = 60$~nm, (c) $x = 0.08$ and $d = 60$~nm, and (d) $x = 0.08$ and $d = 135$~nm. The maximum magnetic field of $\pm 2$~T was applied along the in-plane direction of the devices.}
\end{figure}

The similar hysteresis has often been observed in inhomogeneous superconductivitng materials such as granular cuprate superconductors~\cite{ybco,ybco2,ybco3,ybco4,rusr}, and can be explained by the two-level critical-state model~\cite{tinkham}. In this model, there are two regions, i.e., weak and strong superconducting regions, corresponding to the gray and blue areas in Fig.~2(c), respectively. By applying the magnetic field more than the lower critical field in the positive sweep, fluxons are induced mainly in the weak superconducting region. At a certain magnetic field ($\mu_{0}H_{\rm c}$ in Fig.~3(a)), the fluxons can freely move in the weak superconducting region with a small bias current, resulting in non-zero resistivity at this magnetic field. With increasing the magnetic field more than $\mu_{0}H_{\rm c}$, the fluxons penetrate also into the strong superconducting region and eventually the resistivity becomes back to that of the normal state. We then decrease the magnetic field from the large value. The fluxons start to escape from the weak superconducting region because of the weaker pinning potential compared to the strong superconducting region. As a result, even at the same external magnetic field, there are more fluxons left in the strong superconducting region in the negative sweep. The fluxons trapped in the strong pinning potential cannot move in the sample. This results in a lower resistivity, that is, a higher $\mu_{0}H_{\rm c2}$ ($\mu_{0}H_{\rm r}$ in Fig.~3(a)) in the negative sweep, as schematically illustrated in Fig.~2(c).

In Pt(Bi$_{1-x}$Se$_x$)$_2$ thin-film devices, owing to the spatially non-uniform Se-doping and strain, there should be different regions of different pinning strength with a size of several tens nanometers when it becomes superconducting. As already discussed in Fig.~1(c), the fact that the superconducting transition becomes broader with decreasing $d$ is supportive to apply the two-level critical-state model to our experimental results. We will discuss the origin of the two regions, i.e., weak and strong superconducting regions, in the last part of this section.

To ensure the applicability of the two-level critical-state model to our Pt(Bi$_{1-x}$Se$_x$)$_2$ thin-film devices, we changed several parameters such as the magnetic-field sweep rate, the maximum magnetic field, and the temperature, to measure the hysteresis. Since the magnitude of the hysteresis is larger in the in-plane configuration as shown in Figs.~2(a) and 2(b),  we mainly focus on the magnetoresistance under $\mu_{0}H_{||}$ in the following analysis. In order to quantitatively evaluate the magnitude of the hysteresis in the $\rho$ versus $\mu_{0}H_{||}$ curves, we have defined the difference between $\mu_{0}H_{\rm r}$ and $\mu_{0}H_{\rm c}$ as $\Delta\mu_{0}H\rm{_{c2}} (\equiv $$\mu_{0}H\rm{_{r}} - $$\mu_{0}H\rm{_{c}})$, where $\mu_{0}H\rm{_{r}}$ and $\mu_{0}H\rm{_{c}}$ are the magnetic fields at which $\rho$ takes half of the resistivity in the normal state during the negative and positive sweeps, respectively [see Fig.~3(a)]. 

We first changed the magnetic-field sweep rate, keeping other parameters unchanged. Due to the specification of our current machine, we need to stop the sweeping of magnetic field when taking the resistivity data. Therefore, the magnetic-field sweep rate is defined as the averaged value during the measurement. As shown in Fig.~3(b), $\mu_{0}H\rm{_{r}}$ is almost independent of the magnetic-field sweep rate, while $\mu_{0}H\rm{_{c}}$ decreases as increasing the magnetic-field sweep rate. This results in an increase of $\Delta\mu_{0}H\rm{_{c2}}$ with increasing the sweep rate. The same tendency has been confirmed for multiple devices with the other concentration $x=0.08$ and other thicknesses [see Figs.~3(c) and 3(d)]. The above behavior can be understood within the framework of the two-level critical-state model. As explained in Fig.~2(c), fluxons are first induced in the weak superconducting region as the magnetic field is increased. The induced fluxons move due to a bias electric current and/or thermal fluctuation. In such a process, some fluxons may go into the strong superconducting region and become trapped by a strong pinning potential. If the magnetic-field sweep rate is extremely slow, most of the fluxons are trapped in the strong superconducting region resulting in a lower resistivity and thus higher $\mu_{0}H\rm{_{c}}$. This is more noticeable in the positive sweep because there are more fluxons in the weak superconductivity region, compared to the negative sweep. As a result, $\mu_{0}H\rm{_{c}}$ increases with decreasing the magnetic-filed sweep rate, while $\mu_{0}H\rm{_{r}}$ is almost independent of the magnetic-field sweep rate, leading to the reduction of $\Delta\mu_{0}H\rm{_{c2}}$. 

\begin{figure}
\includegraphics[width=60mm]{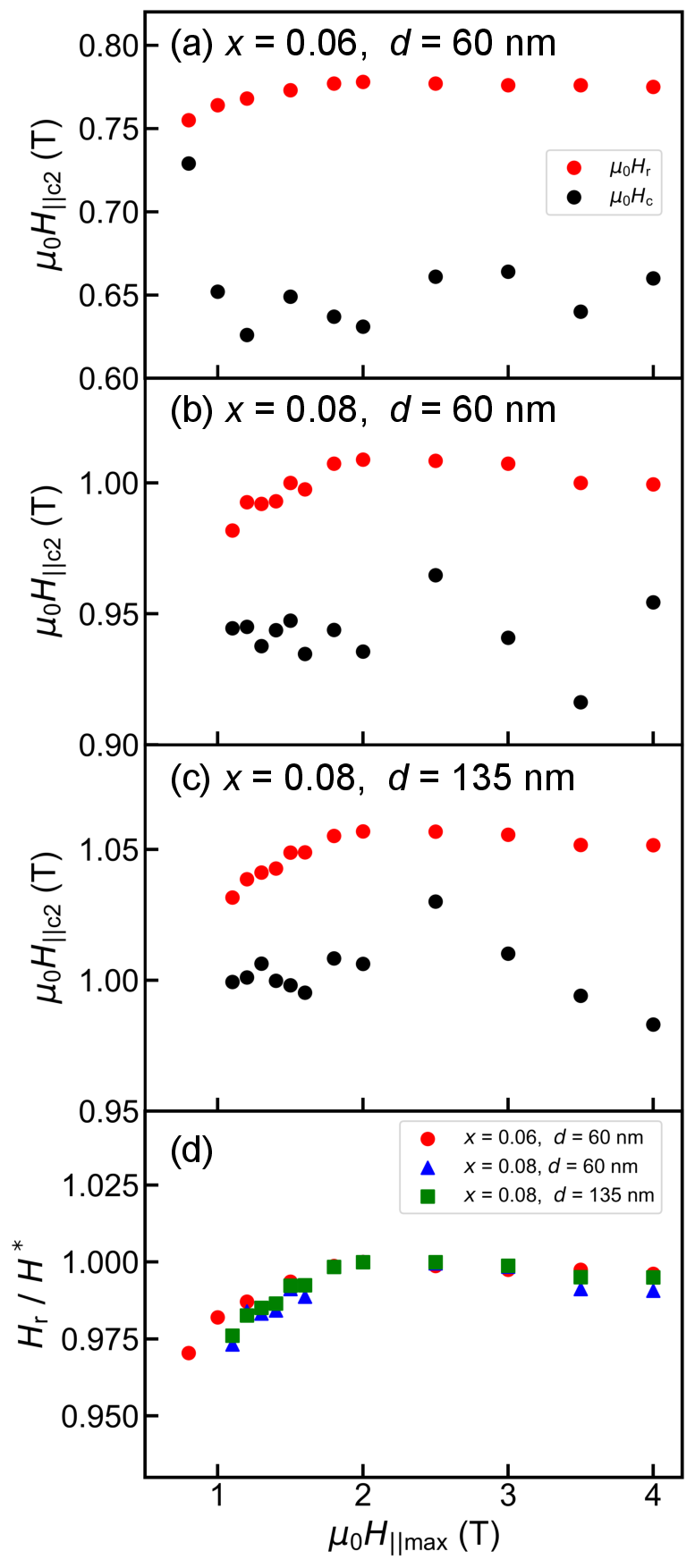}
\caption{\label{fig:epsart} (a)-(c) The maximum in-plane magnetic field $\mu_{0}H_{||\rm{max}}$ dependence of $\mu_{0}H\rm{_{r}}$ (red dots) and $\mu_{0}H\rm{_{c}}$ (black dots) measured at $T = 100$~mK using the same devices as in Figs.~3(b)-3(d). The magnetic-field sweep rate is $0.065$~T/min for (a), and $0.077$~T/min for (b) and (c). (d) The $\mu_{0}H_{||\rm{max}}$ dependence of $\mu_{0}H\rm{_{r}}$ normalized by $\mu_{0}H\rm{^{*}}$, which is the $\mu_{0}H\rm{_{r}}$ value at $\mu_{0}H_{||\rm{max}}=2$~T, obtained with the same devices as in Figs.~3(b)-3(d): $x = 0.06$ and $d = 60$~nm (red circles), $x = 0.08$ and $d = 60$~nm (blue triangles), and $x = 0.08$ and $d = 135$~nm (green rectangles).}
\end{figure}

We next changed the maximum magnetic field, keeping the magnetic-field sweep rate constant ($0.065$~T/min for $x = 0.06$ device and $0.077$~T/min for $x=0.08$ devices). In Figs.~4(a)-4(c), we show the maximum in-plane magnetic filed $\mu_{0}H_{||\rm{max}}$ dependence of $\mu_{0}H\rm{_{r}}$ and $\mu_{0}H\rm{_{c}}$ measured with the same devices as in Fig.~3(b)-3(d). $\mu_{0}H_{\rm{r}}$ gradually increases with increasing $\mu_{0}H_{||\rm{max}}$ and is saturated above 2~T, while $\mu_{0}H_{\rm{c}}$ dose not show a noticeable field dependence. The variation for $\mu_{0}H_{\rm{c}}$ stems from the resistivity jump due to sudden flux changes inside the superconductor, which is also discussed in the previous research~\cite{bini}. To compare $\mu_{0}H\rm{_{r}}$ obtained with different devices, we plot $\mu_{0}H_{\rm{r}}$ normalized by the value of $\mu_{0}H_{\rm{r}}$ at $\mu_{0}H_{||\rm{max}} = 2$~T ($\equiv\mu_{0}H^{\rm{*}}$). For all the investigated devices, it shows the same tendency as explained above. The $\mu_{0}H_{||\rm{max}}$ dependence of $\mu_{0}H\rm{_{r}}$ can also be understood by the two-level critical-state model. Considering the negative sweep process, the system shows zero resistivity when there is almost no fluxon in the weak superconducting region (see the bottom schematic in Fig.~2(c)). In this situation, the external magnetic field is nearly equal to the summation of fluxons trapped in the strong superconducting region. Thus, the magnitude of $\mu_{0}H_{\rm{r}}$ is proportional to the number of fluxons pinned in the strong superconducting region. Even after $\rho$ gets back to the same value as in the normal state under a magnetic field, not all the pinning sites in the strong superconducting region are fully occupied when $\mu_{0}H_{||\rm{max}}$ is not sufficiently large. With increasing $\mu_{0}H_{||\rm{max}}$, we can pin more fluxons in the strong superconducting region. As a result, $\mu_{0}H\rm{_{r}}$ continuously increases until all the pinning sites in the strong superconducting region are occupied, and is eventually saturated. This scenario not only supports the two-level critical-state model but also indicates that the pinning sites are fully occupied at around 2~T in our Pt(Bi$_{1-x}$Se${_x}$)$_2$ thin-film devices. The similar maximum magnetic field dependence of $\mu_{0}H\rm{_{r}}$ was also reported in granular cuprate superconductors~\cite{ybco4} and oxygen-annealed FeTe thin-films~\cite{fete} where all the results were also well-explained by the two-level critical-state model.

\begin{figure}
\includegraphics[width=60mm]{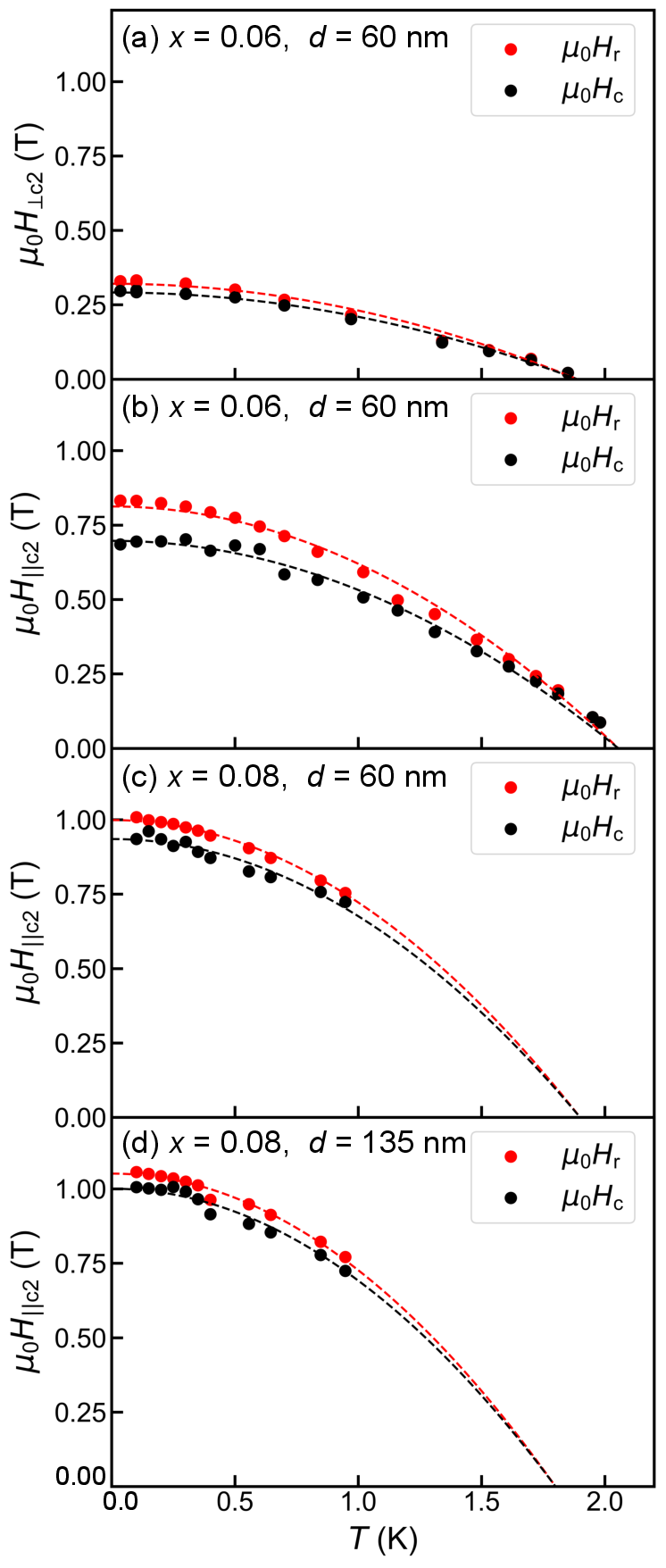}
\caption{\label{fig:epsart} (a), (b) Temperature dependence of $\mu_{0}H\rm{_{r}}$ (red circles) and $\mu_{0}H\rm{_{c}}$ (black circles) under (a) the out-of-plane and (b) in-plane magnetic fields measured with the same device as in Fig.~3(b). (c), (d) Temperature dependence of $\mu_{0}H\rm{_{r}}$ and $\mu_{0}H\rm{_{c}}$ under the in-plane magnetic field using the same devices as in Figs.~3(c) and (d), respectively.
The magnetic-field sweep rates of $0.065$~T/min for (a) and (b), and $0.077$~T/min for (c) and (d). The the maximum filed $\mu_{0}H_{\rm{max}}$ is fixed to 2~T. The dotted lines are the best fits of $\mu_{0}H_{\rm{c2}}(T) = \mu_{0}H_{\rm{c2}}(0)[1-(\frac{T}{T_{\rm{c}}})^2]$. }
\end{figure}

In Fig.~5, we show the temperature dependence of $\mu_{0}H_{\rm{r}}$ and $\mu_{0}H_{\rm{c}}$ measured with the same devices as in Figs.~3 and 4 under the magnetic-field sweep rate of $0.065$~T/min for Figs.~5(a), 5(b) and $0.077$~T/min for Figs.~5(c), 5(d). The maximum field $\mu_{0}H_{\rm{max}}$ has been fixed to 2~T. With increasing temperature, $\mu_{0}H_{\rm{r}}$ and $\mu_{0}H_{\rm{c}}$ monotonically decrease toward $T_{\rm{c}}$, indicating that $\Delta\mu_{0}H_{\rm{c2}}$ becomes zero at $T_{\rm{c}}$. This temperature dependence clearly demonstrates that the observed hysteresis comes from the superconducting properties. The result also supports that the hysteresis is well-explained by the two-level critical-state model. The similar behavior was also reported in FeTe thin films where the two-level critical-state model was used to explain the results~\cite{fete}.

Lastly, we discuss the origin of the two-level critical-state. As already mentioned, a similar hysteresis described by the two-level critical-state model was reported in granular cuprate superconductors~\cite{ybco,ybco2,ybco3,ybco4,rusr}, Bi nanowire~\cite{bi}, FeTe thin-films annealed in oxygen atmosphere~\cite{fete}, and Bi/Ni bilayer~\cite{bini}. In these cases, the origin of the two-level critical-state can be attributed to the granular nature of the amount of oxygen~\cite{ybco,ybco2,ybco3,ybco4,rusr} for cuprate superconductors, the inhomogeneity of oxidized region in FeTe films~\cite{fete}, the spatial thickness fluctuation of ferromagnetic thin film in Bi/Ni bilayer~\cite{bini}.
In our Pt(Bi$_{1-x}$Se$_x$)$_2$ thin-film devices, the inhomogeneity of superconductivity could be induced by the two possible reasons: one is due to spatially non-uniform doping of Se and the other is spatially non-uniform strain.
As explained in Sec.~I, $T_{\rm{c}}$ monotonically decreases as the amount of Se doping is increased after the structural phase transition~\cite{takaki}. This means that the superconductivity is weaker in Se-rich regions, and thus spatially non-uniform doping of Se results in a non-uniform distribution of regions with different superconducting strengths. In addition, strain can also modulate the superconductivity strength~\cite{PtBi2Press}.
As explained in Sec.~II, quenching in ice water was performed to obtain Pt(Bi$_{1-x}$Se$_x$)$_2$ bulk samples. In this process, spatially non-uniform strain would be introduced, because such a rapid temperature change induces spatially non-uniform thermal contraction.
This type of strain induced by high-rate thermal-cycles was also discussed in PtSe$_2$~\cite{PtSe2frag}. Replacement of the Bi sites by Se can also cause local strain because Se has a smaller atomic radius than Bi.
At the moment, however, we cannot conclude which mechanism is more dominant for the present results. Thus, further experiments, such as atomic-scale microscopy and the exploration of various doping materials, are needed to elucidate the origin of the two-level critical-state.

\section{\label{sec:level1}conclusion}
In conclusion, we have measured magnetotransport properties in a vdW superconductor Pt(Bi$_{1-x}$Se$_x$)$_2$. A clear hysteresis behavior was observed in the magnetic field dependence of the resistivity. This hysteresis has the opposite sign to the conventional one. In order to identify the origin of the hysteresis, we varied the magnetic-field sweep rate, the maximum magnetic field, and the temperature when the resistivity was measured as a function of the magnetic field. Through these measurements, we have attributed the origin of such a hysteresis to the two-level critical-state model. The present results demonstrate the importance of the two-level critical-state in vdW superconductors and suggest a new analytical tool for superconducting compounds with large inhomogeneties.

\section{\label{sec:level1}Acknowledgement}
This work was supported by JSPS KAKENHI(Grant Nos. JP20H02557, JP21J20477, JP22H04481, JP23H00257, JP23K13062, JP19H05823 and JP22H01182) and JST FOREST (Grant No.JPMJFR2134).

%\bibliography{Citations}% Produces the bibliography via BibTeX.


\end{document}